\documentclass[12pt]{article}
\usepackage{epsfig}
\usepackage{amsmath}
\usepackage{amssymb}
\normalsize \baselineskip 16pt
\newcommand {\beq} {\begin{equation}}
\newcommand {\eeq} {\end{equation}}
\newcommand {\eps} {\varepsilon}
\newcommand {\gam} {\gamma}

\newcommand {\lam} {\lambda}
\newcommand {\lbr} {\lbrack\!\lbrack}
\newcommand {\rbr} {\rbrack\!\rbrack}

\begin{document}
\title{Beyond kinetic relations}
\author{Lev Truskinovsky \thanks{Laboratoire de Mechanique des Solides, CNRS-UMR 7649,
 Ecole Polytechnique,
91128, Palaiseau, France} \and Anna Vainchtein \thanks{Department
of Mathematics, University of Pittsburgh, Pittsburgh, PA 15260}}
\maketitle

\begin{abstract}

We introduce the concept of
\emph{kinetic equations} representing a natural extension
of the more conventional notion of a
\emph{kinetic relation}. Algebraic kinetic relations, widely used to model
dynamics of dislocations, cracks and phase boundaries, link the instantaneous value of the velocity of
a defect with an instantaneous value of the driving force.
The new approach generalizes kinetic relations by implying a relation
between the velocity and the driving force which is nonlocal in time. To make this relations explicit one needs to
integrate the system of kinetic equations. We illustrate the
difference between kinetic relation and kinetic equations by
working out in full detail a prototypical model of an overdamped defect in a
one-dimensional discrete lattice.  We show that the
minimal nonlocal kinetic description containing now an internal time scale is furnished by a system of two ordinary differential
equations coupling the spatial location of  defect with another
internal parameter that describes configuration of the core region.

\end{abstract}


\section{Introduction}

Kinetic relations attributing a particular value of velocity to
a given value of the driving force are widely used in continuum
mechanics as a constitutive description of such lattice
defects as phase boundaries, dislocations and  cracks  (see
the reviews \cite{AK06, Gurtin99, LeFloch02, Maugin93}). These
algebraic relations form independent postulates that serve as
closing conditions specifying singular solutions in the classical continuum theories.  Since kinetic relations replace
the detailed modeling of the core regions of the defects, they represent
 a condensed description of the complex physical
behavior at the microscale. In practice, kinetic relations are
either taken from experiment or deduced from the solutions of
auxiliary  microscale problems. Characteristically, these auxiliary problems always assume constant values of both the macroscopic velocity
of the defect and of the corresponding driving force.

Due to the implicit assumption that the defect is in a steady motion, the kinetic relations-based description misses the details of a nonsteady internal dynamics of the core region.
In particular, the internal pulsations originating from the defect interaction with localized micro-inhomogeneities become averaged out. This is in contradiction
with the presence of accelerated motions of defects at all scales, as revealed, for instance, by the power law acoustic emission accompanying plasticity, martensitic phase transitions and fracture (e.g. \cite{Kardar98}).

To partially recover the missing information we propose in this paper to replace
algebraic kinetic relations by  differential \emph{kinetic equations}. The aim of these equations is to capture the transient phases of the defect evolution
in response to nonsteady driving. In the language of constitutive theory we propose to replace the instantaneous
rheological relations on the phase boundary by nonlocal memory functionals originating from a local
description in terms of internal variables. Such extension brings into the conventional theory an internal time scale and allows one to deal with the
so-called ``rate effects".
An example of a similar development is provided by the rate-dependent constitutive
laws in the theory of friction, where the set of internal state
variables is also assumed to satisfy differential constitutive
relations (e.g. \cite{RR83,Ruina}).

In order to show the possibility of a systematic derivation of kinetic equations from a micromodel
we consider a prototypical overdamped defect moving in a lattice. We develop a low-parametric description
of the internal dynamics of this defect
 involving
some specially selected internal variables which characterize the structure of the core region.
Our approach can be viewed as an example of a quasicontinuum method in the sense of \cite{AL08, LiE05, TOP96} whose goal is to match the
macroscopic continuum description outside singularities with a
more detailed atomistic resolution of the core regions. The zero-order model of this type produces kinetic relations and implies instantaneous and
universal response of the core region to external perturbations (autonomous core region).
The first-order approximation leads  to kinetic equations which already capture some nonuniversal features of the microscopic dynamics.
 For consistency of the two
approximations, kinetic equations must of course reduce to the kinetic relation when the variation of the macroscopic parameters is
sufficiently slow.

To illustrate the main idea of our approach consider a toy model describing an overdamped  dynamics of a
configurational point in a one-dimensional energy landscape:
\begin{equation}
\label{eq:kin_eqn}
\frac{\partial \nu}{\partial \tau}=-
\frac{\partial \Phi(\nu;G(t))}{\partial \nu}.
\end{equation}
Here $\nu$ is a variable defining the microstate of the system and
$G(t)$ is a sufficiently slow varying macroscopic driving force which depends
on slow time $t$. The fast time is defined as $\tau=t/\epsilon$,
where $\epsilon<<1$ is a small parameter. We assume that the gradient of the
energy landscape $\Phi(\nu;G)$ is periodic in $\nu$. The method of kinetic relations postulates the existence of an
 algebraic relation between the
driving force $G$ and the macroscopic velocity
\begin{equation}
V(t)=\biggl\langle\frac{\partial \nu}{\partial
\tau}\biggr\rangle_{\tau}=\lim_{\epsilon\rightarrow0}\epsilon\int_{0}^{1/\epsilon}\frac{\partial\nu}{\partial
\tau}d\tau.
\label{eq:V_average}
\end{equation}
Such relation, which we write as  $V(t)=\Gamma (G(t))$, can be computed explicitly if $G$
is a tilt of the energy landscape
\begin{equation}
\Phi(\nu;G)=G_{\rm P}\cos\nu-G\nu. \label{eq:wiggly_energy}
\end{equation}
Then a direct computation gives (e.g. \cite{ GZC81,ACJ96})
\begin{equation}
V = \Gamma(G) =
\begin{cases}
0 & |G| \leq G_{\rm P}\\[6pt]
\;{\rm sgn}(G-G_{\rm P})\sqrt{G^2-G_{\rm P}^2}& |G|\geq G_{\rm P}.
\label{eq:wiggly_energy1}
\end{cases}
\end{equation}
In the realistic situations the microscopic description of the
type \eqref{eq:kin_eqn} is too complex because it involves a huge
number of variables. In contrast, the
macroscopic description
\eqref{eq:wiggly_energy1} is too schematic and cannot be trusted
when one deals with the problems where slow and fast time scales
cannot be separated. This point has been often overlooked, and kinetic
relations implying a universal elimination of microscopic
variables were used in situations where both $G$ and $V$ are
changing fast as, for instance, in the case of pinning-depinning
phenomena.

In what follows we present a detailed adaptation of the above ideas to
the case of a  martensitic phase transition.  Martensitic phase
boundaries are particularly convenient for the demonstration of
the main principles of our approach
 because these plane defects may be
adequately represented already in one-dimensional models. To
emphasize ideas we consider the simplest case of a phase
boundary with overdamped dynamics. At the microscale, the analysis
of the non-steady evolution of the core region of such phase boundaries
requires a study of a dynamical system with an infinite number of
degrees of freedom. At the macroscale the phase transition is
modeled as a singular surface (jump discontinuity) whose evolution is
governed by a kinetic relation \cite{Trusk82}.

We pose the question of whether an intermediate description is
possible when the interface is equipped with a small number of
``mesoscopic'' degrees of freedom whose dynamics reproduces the
main transient effects. An early  example of such a reduced description can be found in
the theory of dislocations where
static defects are often represented as  effective particles in the Peierls-Nabarro
(PN) landscape \cite{BraunKivshar04}. The static PN landscape is
obtained by relaxing all microscopic variables other than one
 collective variable interpreted as the macroscopically observable
location of the core. The idea of a tilted PN landscape has been heuristically applied
to the description of dynamic dislocations in close to  continuum limit \cite{FlachKladko96, Hobart65, Hobart66,
IshibashiSuzuki84, IshimoriMunakata82, Lazutkin89, Pokrovsky81, Willis86}. This approach, however, cannot be used in principle in the
strongly discrete case when the  dislocation core is atomically narrow \cite{Furuya87,Joos82}. In order to deal with this limit
it is natural to abandon the idea of macroscopic collective variables and
 trace instead the dynamics of the particular discrete elements while again enslaving all others.  Such elements were called ``localized normal modes''
in \cite{W64} and ``active points'' in \cite{KMB00}. In some special cases it has been proven rigorously that dynamics of ``active points'' corresponds
to the motion along the center manifold of the infinite-dimensional dynamical system \cite{CB2}.

In this paper we extend the approach of ``active points''
from the immediate vicinity of the depinning
point where it has been formally justified to a broader class of nonsteady motions of defects.  The key to our
method is the assumption that the dynamics of only a few bonds located in the core region 
has to be resolved fully. The other bonds remain confined to near bottoms of their respective potential
wells, and their small adjustment to the changing conditions can be
treated as instantaneous.

The main question  confronted by such theory concerns the minimal number of internal variables
which is sufficient to capture the response of a defect to a particular class of external perturbations.
In this paper we propose to base the  selection of the minimal set of internal variables on the careful analysis of the relative ``activity"
of the variables in the traveling wave solution. As a result, we obtain kinetic equations which describe faithfully only small deviations
from such steady states. Strong deviations from the traveling wave ansatz have been rigorously studied for several classes of nonlinear equations
in the case of periodic obstacles (pulsating traveling fronts, e.g. \cite{BH02,DY06}). Our approach deals with the singular limit of such problems
(localization of the fronts), which, as far as we know, has not yet been studied in the mathematical literature.

In our construction we use essentially the fact that the exact traveling wave solution of our discrete problem is known in the case
of a piecewise quadratic interaction potential \cite{TV3a, TV5}. Using this solution as a
benchmark, we construct a simple two-dimensional dynamical system which in
the case of a constant macroscopic driving force generates a
remarkably good approximation for the kinetic relation originating from the study of the  full
infinite-dimensional dynamical system. We then apply this low-parametric description to the case when the driving force
is a given function of time and show how it can be used to obtain a rheological model of the transformation kinetics which is nonlocal in time.
We also provide an example showing that depending on the rate of external driving the responses of the
phase boundary based on the kinetic relation and on the associated kinetic equations can be markedly different.
A systematic application of the obtained
kinetic equations to solutions of the particular
macroscopic boundary value problems involving strongly
inhomogeneous media will be given elsewhere.

The paper is organized as follows. In Section 2 we formulate the
singular macroscopic problem which requires a microscopic closure.
In search for such a closure we turn in Section 3 to a specific microscopic model.
To illustrate the existing methods of model reduction in relation to moving defects, we construct in  Section 4
the static PN landscape and then use its tilted version in Section 5 to describe the steady state dynamics of phase boundaries.
Since this approach is not satisfactory, we develop in Section 6 a more systematic one-parametric kinetic equation and introduce a concept of dynamic PN landscape.
We then find that the one-parametric model is
only adequate in the immediate
vicinity of the depinning point and proceed in Section 7 with the systematic derivation of the
two-dimensional system of kinetic equations. We find that such description works well
in the whole interval of admissible velocities. In Section 8 we choose a specific
loading program and compare the predictions of the nonlocal model with the predictions based on the
classical kinetic relations. Our concluding remarks are collected in Section 9. To make the exposition self-contained,
we present in the Appendix a concise derivation of the exact kinetic relation.

\section{Macroscopic model }

Let $u(x,t)$ be the one-dimensional continuum displacement field.
The macroscopic energy of a bar undergoing martensitic phase
transition can be written as
\beq
\mathcal{E}=\int\biggl[\dfrac{\rho u_t^2}{2}+\phi (u_{x})\biggr]dx.
\label{eq:energy0}
\eeq
Here $u_{t} \equiv \partial u/\partial
t$ is the macroscopic velocity, $u_x \equiv \partial u/\partial
x$ is the macroscopic strain, $\rho>0$ is the reference mass density,
and the energy density $\phi (u_{x})$ is
represented by a double-well potential which will be specified
later.

The dynamic equation corresponding to the energy
(\ref{eq:energy0}) is
\beq
\rho u_{tt}=(\hat{\sigma}(u_x))_x,
\label{eq:continuum}
\eeq
where $\hat{\sigma}(u_x)= \phi'(u_x)$ is the stress-strain relation.
On discontinuities,  the equation \eqref{eq:continuum} must be
supplemented with the jump conditions. Let $f_{-}$ and
$f_{+}$ denote the values of $f(x)$ to the left and to the right
of the interface, and introduce the notations $\lbr f \rbr \equiv
f_+ -f_-$ for the jump and $\{f\} \equiv (f_{+}+f_{-})/2$ for the
average of $f$ across the discontinuity. Then the parameters on a
discontinuity must satisfy the classical Rankine-Hugoniot jump
conditions
\beq
\lbr u_t \rbr +V\lbr u_x \rbr=0, \quad
\rho V\lbr u_t \rbr+\lbr \hat{\sigma} (u_x)\rbr=0.
\label{eq:jump_conds}
\eeq
The entropy inequality, which must also
hold on the jump, can be written in the form \beq \mathcal{R}=GV \geq 0,
\label{eq:entropy} \eeq where $\mathcal{R}$ is the rate of energy
dissipation and
\beq
G=\lbr \phi \rbr - \{\hat{\sigma}(u_x)\}\lbr
u_x \rbr
\label{eq:G_cont}
\eeq
is the configurational (driving) force. In one-dimensional problems imposing the entropy inequality can remedy local nonuniqueness
only in the case  of supersonic discontinuities (shock waves).
 For subsonic phase boundaries one must specify the rate of entropy production as well.
This is usually done phenomenologically through the kinetic
relation $G=G(V)$ between the driving force and the velocity
\cite{AK91a, Trusk82, Trusk87}. If the microscopic model
 is known one can instead compute the function $G(V)$ directly from the
microscopic traveling wave solution (e.g. \cite{Slemrod83,SCC05,Trusk82,TV3a,TV5}).

Behind the assumption that there exists an algebraic relation between the driving force $G$ and the velocity $V$
 lies the idea that the width of the transition layer is so small that the
internal relaxation to the traveling wave profile takes place much faster that the variation of the driving force. However,
the convergence to the sharp interface limit is expected to be nonuniform with respect to the parameter describing the rate of configurational loading. For instance, if the phase boundary interacts with an obstacle, the structure of
this obstacle should be sufficiently diffuse for the algebraic kinetic relation to hold within certain error bars. If the profile of the obstacle is sufficiently sharp,
the time scales of internal relaxation and of external driving are comparable and the
traveling wave profile is not an adequate ansatz. In this case one can expect a uniform double-parametric asymptotics in the spirit of \cite{BT08} to be more complex than is assumed by
the computations based on the traveling waves.

In what follows we do not attempt a rigorous asymptotic analysis in the general situation and instead choose a simple microscopic
model and formally derive a set of differential kinetic equations describing the transient response for the inner structure of a phase
boundary.

\section{Microscopic model }

\begin{figure}
\centerline{\psfig{figure=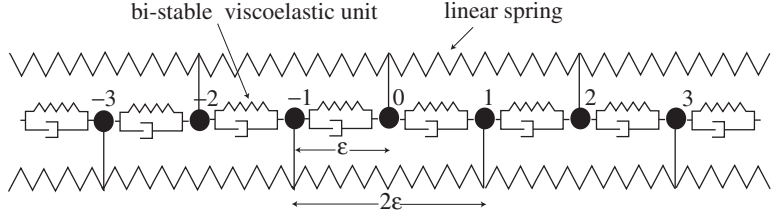,width=4.in}} \caption{The
discrete microstructure with viscoelastic nearest and elastic
next-to-nearest-neighbor interactions.}
\label{fig:infinite_chain_VE}
\end{figure}

Consider an infinite chain of particles,
connected to their nearest neighbors (NN) through viscoelastic
springs and to their next-to-nearest neighbors (NNN) by elastic
springs (see Figure~\ref{fig:infinite_chain_VE}).
Suppose that in
the undeformed configuration the NN and NNN springs have lengths
$\eps$ and 2$\eps$, respectively.  Let $u_n(t)$ denote the
displacement of $n$th particle at time $t$ with respect to the
reference configuration. We associate with the deformation of $n$th NN spring a discrete measure of strain
\beq
w_n=\dfrac{u_{n}-u_{n-1}}{\eps}.
\label{eq:wk_def}
\eeq
For the viscoelastic NN springs we assume
the following constitutive relation: \beq f_{\rm
NN}(w,\dot{w})=\phi_{\rm NN}'(w)+\xi \dot{w}, \label{eq:fNN} \eeq
where $\xi>0$ is the viscosity coefficient. To describe
domain boundaries the function $\phi_{\rm NN}(w)$ must be at least a
double-well potential; to obtain explicit solutions, we
assume that this function is biquadratic:
\beq
\phi_{\rm
NN}(w)=\left\{\begin{array}{ll}
                          \frac{1}{2}\mu w^2, & w \leq w_c\\
                          \frac{1}{2}\mu (w-a)^2
                      +\mu a\biggl(w_c-\dfrac{a}{2}\biggr), & w \geq w_c.
                                            \end{array}
\right.
\label{eq:first}
\eeq
Under these
assumptions the NN elastic units can be in two different phases,
depending on whether the strain is below (phase I) or above (phase
II) the critical value $w_c$. The elastic modulus in each phase is $\mu>0$, and the parameter $a>0$ measures the
transformation strain.  To simplify calculations, we assume
that the
NNN interactions are linearly elastic: \beq f_{\rm NNN}(\hat
w)=2\gam \hat w. \label{eq:fNNN} \eeq Here  we defined $\hat
w_n=(w_{n+1}+w_n)/2$ as the strain in the NNN spring
connecting $(n+1)$th and $(n-1)$th particles.

The dynamics of the chain is governed by the following system of
ordinary differential equations:
\beq
\begin{split}
\rho\eps\ddot{u}_n=& \mu [w_{n+1}-w_n-\theta(w_{n+1}-w_c)a+\theta(w_n-w_c)a]\\
&+\gam(w_{n+2}+w_{n+1}-w_n-w_{n-1})+\xi(\dot{w}_{n+1}-\dot{w}_n).
\end{split}
\label{eq:full_dynamics}
\eeq
Here $\rho>0$ is the mass density of the chain and
$\theta(x)$ is the unit step function.
To ensure stability of the
chain we  require that \beq
E=\mu+4\gam>0, \label{eq:E} \eeq where $E$ is the homogenized
macroscopic elastic modulus; following \cite{TV2,TV1}, we also
assume that the NNN interactions are of ferromagnetic type,
meaning that $\gam \leq 0$. In the limit $\eps/L \rightarrow 0$,
where $L$ is a macroscopic length scale, we recover \eqref{eq:continuum}
with the homogenized stress-strain law \cite{TV5}
\[
\hat{\sigma}(w)=E(w-\Delta\theta(w-w_c)).
\]
Here $\Delta$ is the macroscopic transformation strain:
\beq
\Delta=\dfrac{a \mu}{E}.
\label{eq:Delta}
\eeq
There are two time scales associated with this problem: the time
scale of inertia, $T_{\rm in}=\eps\sqrt{\rho/E}$, and the
viscosity time scale, $T_{\rm visc}=\xi/E$. In this paper we
consider the overdamped limit when $T_{\rm visc}>>T_{\rm in}$,
i.e. \beq \xi>>\eps\sqrt{\rho E}. \label{eq:assumption} \eeq We
can nondimensionalize the problem using $T_{\rm visc}$ as the time scale
and letting
\beq
\bar{t}=\dfrac{tE}{\xi},\quad
\bar{u}_n=\dfrac{u_n}{\Delta\eps}, \quad \bar{w}_n=\dfrac{w_n}{\Delta},
\quad \bar{w}_c=\dfrac{w_c}{\Delta}.
\label{eq:nondim_vars}
\eeq
Dropping the bars on
the new variables, we obtain  dimensionless system equations:
\beq
\dot{w}_n-\dot{w}_{n+1}=\hat{\sigma}(w_{n+1})-\hat{\sigma}(w_n)
+D(w_{n+2}+w_{n+1}-w_n-w_{n-1}),
\label{eq:rescaled_OD}
\eeq
where
\beq \hat{\sigma}(w)=w-\theta(w-w_c)
\label{eq:sigma}
\eeq
is the rescaled macroscopic stress-strain law. The dimensionless parameter \beq
D=-\dfrac{\gam}{E} \geq 0 \label{eq:beta} \eeq measures the
relative strength of NN and NNN interactions. In what follows it
will also be interpreted as a measure of coupling of the bistable
units.

Observe that system \eqref{eq:rescaled_OD} can be
``integrated'', yielding
\beq
\dot{w}_n=D(w_{n+1}-2w_n+w_{n-1})-\hat{\sigma}(w_n)+\sigma.
\label{eq:main}
\eeq
Here $\sigma=\sigma(t)$ is the time-dependent applied stress.
One can see that \eqref{eq:main} also governs the overdamped
Frenkel-Kontorova model \cite{CB2,Fath98,Keener87,KT07}; a subtle but important
distinction of the present setting is that here the
coupling coefficient $D$ is independent of $\eps$.

In the overdamped limit the macroscale governing equations \eqref{eq:continuum} reduce to
$(\hat{\sigma}(u_x))_x=0$, or $\sigma(u_x)=\sigma(t)$, while the jump
conditions take the form $\lbr \hat{\sigma} (u_x)\rbr=0$ \cite{TV5}. The driving force \eqref{eq:G_cont}
is then a function of time:
\[
G(t)=\sigma(t)-\sigma_{\rm M},
\]
where \beq \sigma_{\rm M}=w_c-1/2 \label{eq:Maxwell} \eeq is the Maxwell stress.
The function $G(t)$ also enters the microscopic equation \eqref{eq:main} which can be rewritten in the gradient-flow
form
\beq
\mathbf{\dot{w}}=-\nabla \mathcal{W}(\mathbf{w};G(t)).
\label{eq:inf_system} \eeq
Here $\mathbf{w} \in
\mathbb{R}^{\infty}$ is the vector of strains, the gradient is
taken with respect to $\mathbf{w}$, and \beq
\mathcal{W}=\sum_{n=-\infty}^{\infty}\biggl(\dfrac{1}{2}w_n^2
-(w_n-w_c)\theta(w_n-w_c)+D(w_{n+1}-w_n)^2-(\sigma_{\rm
M}+G(t))w_n\biggr), \label{eq:Gibbs_energy} \eeq is the dimensionless
energy of the system.

Our main goal will be to approximate the
infinite-dimensional dynamical system \eqref{eq:inf_system} by a
finite-dimensional reduced dynamical system of the type \beq
\dot{\boldsymbol{\nu}}=-\boldsymbol{\alpha}\nabla\Phi(\boldsymbol{\nu};G(t)),
\label{eq:reduced_system} \eeq where $\boldsymbol{\alpha}$ is the
effective mobility matrix. The gradient is taken with respect
to the order parameter $\boldsymbol{\nu} \in \mathbb{R}^{K}$ where the
integer-valued parameter $K$ defines the dimensionality of the
reduced system.
After the solution of the vector equation
\eqref{eq:reduced_system} is known,
 the approximation of the discrete field \eqref{eq:inf_system} should be recoverable from the auxiliary relations
$w_n(t)=w_n(\nu_1(t),\dots,\nu_K(t))$ describing the recovery of relaxed ``non-order-parameter'' variables.

In the absence of a rigorous method  condensing  \eqref{eq:inf_system}
to \eqref{eq:reduced_system} we base our formal reduction on the detailed
study of a traveling wave solution of \eqref{eq:inf_system} in the form $w_n(t)= w(n-Vt)$. This solution, which is known explicitly (see Appendix),
corresponds to the case when $G(t)=\text{const}$ and describes a steadily propagating phase boundary. We emphasize that in our method the $G=\text{const}$
solution is used only to suggest the dimensionality of the reduced system,
while the actual behavior of the resulting dynamical system is fully driven by the time-dependent $G(t)$.

\section{Classical PN landscape}
\label{sec:statics}

The simplest one-parametric ($K=1$) reduction of the infinite-dimensional system \eqref{eq:inf_system} is based on the idea
of Peierls-Nabarro (PN) energy landscape which was first developed for dislocations (see the review \cite{Nabarro}) and then generalized to phase boundaries in \cite{TV2}. The
construction is fully static and therefore restricted to the case of sufficiently small $G$. Since the ideas behind the construction
of the PN landscape are used later to obtain the approximate kinetic equations, we outline the main steps below; see \cite{TV2} for more details.

Assume that $G=\sigma-\sigma_{\rm M}=const$ and compute stable monotone
solutions of the equilibrium difference equations \eqref{eq:inf_system}. For the phase boundary located at $n=m$
we obtain (see also \cite{Fath98,Hobart65}): \beq w_n^m(G)=\sigma_{\rm M}+\left\{
\begin{array}{ll}
              G+1-\dfrac{\exp(\lam(n-m-1/2))}{2\cosh(\lam/2)},
& \mbox{$n<m$} \\
              G+\dfrac{\exp(-\lam(n-m-1/2))}{2\cosh(\lam/2)},
& \mbox{$n \geq m$},
              \end{array}
      \right.
\label{eq:statics} \eeq where \beq \lam (D)={\rm
arccosh}\biggl(1+\dfrac{1}{2D}\biggr). \label{eq:lam} \eeq The
admissibility  constraints \beq w_n^m \geq w_c \;\;\;\;\mbox{\rm
for $n \leq m$},\;\;\;\;\;\; w_n^m \leq w_c\;\;\;\; \mbox{\rm for
$n \geq m+1$} \label{eq:constraints_statics} \eeq determine the
constraints on $G$; the set of driving forces satisfying these
constraints constitutes the \emph{trapping region}. One can show
that for $D>0$ the strain profile (\ref{eq:statics}) is monotone,
so the constraints (\ref{eq:constraints_statics}) can be replaced
by $w_m^m \geq w_c$ and $w_{m+1}^m \leq w_c$. The trapping region
is then given by \beq |G| \leq G_{\rm P}, \label{eq:trap_region}
\eeq where \beq G_{\rm P}(D)=\dfrac{1}{2}\tanh
\dfrac{\lam}{2}=\dfrac{1}{2\sqrt{1+4D}}. \label{eq:FP} \eeq is the
desired expression for the Peierls threshold \cite{BKZ90,TV2}.

One can see that at $|G| \leq G_{\rm P}$  the phase boundary can
be in an infinite number of  stable equilibrium configurations \eqref{eq:statics}
parameterized by $m$.
 In order to connect these stable states one needs to consider a path involving non-equilibrium
intermediate configurations. For instance, take two
equilibrium configurations, one with the phase boundary located at
$n=i-1$ and the other one at $n=i$. Suppose that phase II is located
behind the phase boundary and  consider all paths connecting the two
equilibria along which the $i$th NN spring is the only one that
changes phase, while all other springs stay in their respective
energy wells. As the $i$th NN spring crosses the critical value of
strain $w_c$ the system goes through an energy
barrier and the next task is to select the
path that involves the minimal barrier.

To this end we fix $w_i$ and
minimize the energy of the chain with respect to all strains
$w_k$ with $k \neq i$. We obtain: \beq
(1+2D)w_k-D(w_{k+1}+w_{k-1})=\sigma_{\rm
M}+\left\{\begin{array}{ll}
                                                     G+1, & k \leq i-1\\
                                                     G, & k \geq i+1.
                                        \end{array}
                             \right.
\label{eq:constr_equil} \eeq Solving equations \eqref{eq:constr_equil} at $k \leq i-1$ and $k
\geq i+1$ and requiring that the corresponding solutions, when
extended to $k=i$, both equal $w_i$, we obtain the following strains along the path connecting
the two equilibrium states:\footnote{Notice that along the selected path the order parameter $w_i$ increases
from its value in the first minimum,
$w_{\rm L}=w_i^{i-1}(G)=\sigma_{\rm
M}+G+\exp(-\lam/2)/(2\cosh(\lam/2)) \leq w_c$,
to its value in the second minimum, $w_{\rm U}=w_i^i(G)=\sigma_{\rm
M}+G+1-\exp(-\lam/2)/(2\cosh(\lam/2)) \geq w_c.
$}
\beq
w_k=\sigma_{\rm M}+\left\{\begin{array}{ll}
             G+1+e^{\lam(k-i)}(w_i-\sigma_{\rm M}-G-1), & k \leq i\\
             G+e^{-\lam(k-i)}(w_i-\sigma_{\rm M}-G), & k \geq i.
                                        \end{array}
                             \right.
\label{eq:wk} \eeq
 At the saddle point $w_i=w_c$ the energy reaches its maximal value; due to the
nonsmoothness of the biquadratic potential at $w=w_c$, the corresponding states are singular.

The next step is to construct a global path that connects not just two but
all equivalent equilibrium points. To this end we replace the order
parameters  $w_i$  which were operative in the consecutive
segments of the path by a global order parameter $\nu$ defined
implicitly by \beq w_{[\nu]}=w^*(\nu)=\sigma_{\rm
M}+G+\dfrac{\exp(-\lam/2)}{2\cosh(\lam/2)}
+(\nu-[\nu])\tanh(\lam/2). \label{eq:wstar_below_GP} \eeq Here
$[\nu]$ denotes the integer part of $\nu$. At the integer values of $\nu$ we have
$w_{[\nu]}=w_{\rm L}$, and as $\nu$ approaches $[\nu]+1$ from
below, $w_{[\nu]}$ linearly increases to $w_{\rm U}$. The function
$w^*(\nu)$ is periodic with period $1$ and has jump
discontinuities at $\nu=[\nu]$. One can see that $\nu$ can be
interpreted as the macroscopic location of the interface.
\begin{figure}
\centerline{\psfig{figure=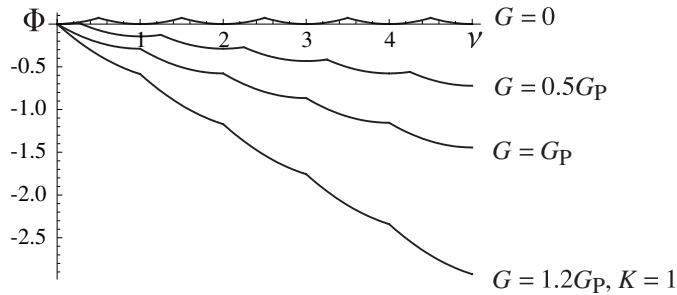,width=3.5in}}
\caption{One-dimensional energy landscape $\Phi(\nu;G)$ at $D=0.5$
and different $G$. The classical static PN landscape corresponds
to $G \leq G_{\rm P}$. The dynamic potential above the Periels
threshold $G > G_{\rm P}$ was constructed using the $K=1$
approximation.} \label{fig:1Dpotential}
\end{figure}

Using (\ref{eq:wstar_below_GP}) and recalling (\ref{eq:wk}), we
obtain strains along the path that are now parametrized by $\nu$: \beq
w_k=\sigma_M+\left\{\begin{array}{ll}
             G+1+e^{\lam(k-[\nu])}(w^*(\nu)-\sigma_{\rm M}-G-1), & k \leq [\nu]\\
             G+e^{-\lam(k-[\nu])}(w^*(\nu)-\sigma_{\rm M}-G), & k \geq [\nu].
                                        \end{array}
                             \right.
\label{eq:path} \eeq To evaluate the energy
$\mathcal{W}$ along the path (\ref{eq:path}) we first renormalized it as
$\Phi(\nu;G)=\mathcal{W}(\nu)-\mathcal{W}(0)$. Substituting
\eqref{eq:path} in \eqref{eq:Gibbs_energy}, we obtain (recall that $0 \leq
G \leq G_{\rm P}$) : \beq \Phi(\nu;G)=G_{\rm P}(\nu-[\nu])^2-G[\nu]
-(G-G_{\rm P}+2(\nu-[\nu])G_{\rm
P})\theta\biggl(\nu-[\nu]-\dfrac{G_{\rm P}-G}{2G_{\rm P}}\biggr).
\label{eq:path_below_GP} \eeq Observe that the minima of the resulting PN landscape $\Phi(\nu;G)$ are
located at the integer values of $\nu$ and correspond to stable
equilibrium states, while the singularities at $\nu=\nu_i$
represent unstable equilibria (saddle points) where $w_i(\nu_i)=w_c$ (see
Fig.~\ref{fig:1Dpotential}). The energy barriers separating  equilibria at $\nu=i$ and
$\nu=i+1$ are independent of $i$:
\[
\Phi(\nu_i)-\Phi(i)=\dfrac{(G_{\rm
P}-G)^2}{4G_{\rm P}}.
\]
At $G=G_{\rm P}$ the stable equilibria become
marginally stable; see the curve $G=G_{\rm P}$ in
Fig.~\ref{fig:1Dpotential}. Above the Peierls threshold ($G_{\rm
P}<G<G_{\rm S}$), the stable equilibria cease to exist, and the above
construction of the PN landscape becomes invalid.

\section{Dynamics on a tilted PN landscape}

One very heuristic but rather common way to extend the idea of
the PN landscape to dynamics is to take the classical PN
landscape at $G=0$ and  tilt it to account for nonzero $G$.
Then one can obtain an approximate kinetic relation by considering the overdamped motion of a material point on this tilted landscape
and studying the dependence of the average velocity on the driving force \cite{BraunKivshar04}.

The tilted PN landscape can be defined as follows \beq
\Phi_{\rm T}(\nu;G)=\Phi(\nu+1/2;0)-G\nu-\dfrac{G_{\rm P}}{4},
\label{eq:tilted} \eeq where we have added a constant to
ensure that $\Phi_{\rm T}(0;G)=0$. It is not hard to see that at
$G \leq G_{\rm P}$ the energy \eqref{eq:path_below_GP} agrees with
\eqref{eq:tilted} up to
 the appropriate shifts in horizontal and vertical directions:
\[
\Phi\biggl(\nu+\dfrac{G_{\rm P}-G}{2G_{\rm
P}};G\biggr)-\dfrac{(G_{\rm P}-G)^2}{4G_{\rm P}}=\Phi_{\rm T}(\nu;
G), \quad 0 \leq G \leq G_{\rm P}.
\]
Note that the two landscapes coincide exactly only at the Peierls threshold. Indeed, in the actual PN landscape \eqref{eq:path_below_GP}
the local energy minima are always located at the integer values
of $\nu$, reflecting the fact that in a discrete system position
of a phase boundary always coincides with a lattice point. As $G$ approaches the value $G_{\rm P}$ the points
of local maxima of the PN landscape also move towards integer values and eventually the minumum and maximum points merge.
Meanwhile, in the case of the tilted PN landscape
\eqref{eq:tilted} the local maxima are always fixed at the integer
values, and the minimum points move instead.
\begin{figure}
\centerline{\psfig{figure=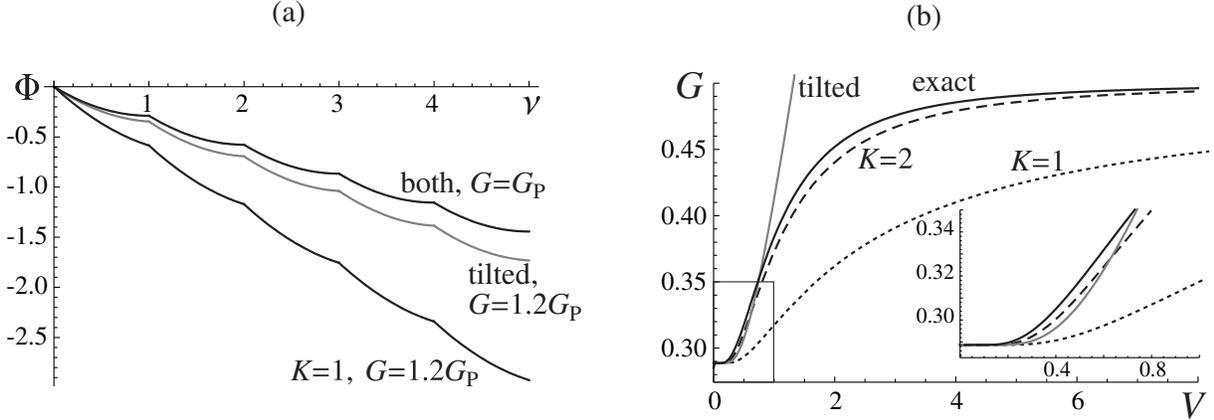,width=\textwidth}} 
\caption{(a) The tilted energy landscape (grey) and the DPN landscape for $K=1$
(black) at $D=0.5$. The two landscapes coincide at $G=G_{\rm P}$.
(b) The exact kinetic relation $G(V)$ from the Appendix (solid curve) and its
approximations via the tilted landscape (grey curve), $K=1$
approximation (dotted curve) and $K=2$ approximation (dashed
curve).} \label{fig:tilted}
\end{figure}

Despite these deficiencies, the tilted PN landscape has a strong advantage: it can be formally  extended beyond the Peierls
threshold $G=G_{\rm P}$. One can then try to model the dynamics of a
phase boundary by the overdamped motion of a particle in the tilted
landscape (see the grey curve in Figure~\ref{fig:tilted}a). The
equation of motion takes the form
\[
\dot{\nu}=-\alpha \Phi'_{\rm T}(\nu;G)=-\alpha(2G_{\rm P}(\nu-[\nu]-1/2)-G).
\]
Solving this equation subject to the condition $\nu([Vt]/V)=[Vt]$
yields
\[
\nu(t)=[Vt]+\dfrac{G+G_{\rm P}}{2G_{\rm P}}\biggl(1-e^{-2\alpha
G_{\rm P}(t-[Vt]/V)}\biggr).
\]
Imposing the periodicity condition $\nu(([Vt]+1)/V)=[Vt]+1$, we obtain the desired kinetic relation
\beq G(V)=G_{\rm P}\dfrac{1+e^{-2\alpha G_{\rm
P}/V}}{1-e^{-2\alpha G_{\rm P}/V}}. \label{eq:G_tilted} \eeq
As expected, it gives the correct limit $G \rightarrow G_{\rm P}$
when $V \rightarrow 0$. However, in contrast to the exact
kinetic relation obtained in the Appendix (see \eqref{eq:G_OD}) the function \eqref{eq:G_tilted}
is unbounded: at large $V$ it grows as $G(V) \approx V/\alpha$ (see
Figure~\ref{fig:tilted}b). In addition, since the physical meaning
of the dynamic variable $\nu(t)$ is obscure, it is not clear how
one can recover the strains $w_n(t)$. We therefore abandon this heuristic path and search for a more adequate representation of the actual dynamics of the chain.

\section{Reduced model with one internal variable}

We now take a more systematic point of view,
choose a single variable and follow its exact dynamics while minimizing the energy with respect to all
remaining variables. The resulting model is similar to the ``single-active-site theory'' of \cite{CB2,KMB00}.

\subsection{$K=1$ model}

Suppose that  $G=\text{const}$ and assume that the motion in the reduced system is periodic. Assume also that during
each period dynamics of only a single NN spring, located right behind the
phase transition front and actually changing the energy well
during this period, needs to be traced in all the detail, while all other springs can be ``enslaved''.
We can always assume that at $t=0$ the $0$th NN spring has just switched
to phase II: $w_0(0)=w_c$. Therefore, during the time period $0
\leq t < 1/V$ the ``active strain'' is  $w_0$. Minimizing the
energy with respect to all other strain variables, we obtain \beq
w_n=w_c-1/2+G+\left\{\begin{array}{ll}
             1+(w_0-w_c-G-1/2)e^{\lam n}, & n \leq 0\\
             (w_0-w_c-G+1/2)e^{-\lam n} & n \geq 1.
                                        \end{array}
                             \right.
\label{eq:other_strains_1D} \eeq Substituting these expressions in
\eqref{eq:main} for $n=0$, we obtain a single equation governing the
dynamics of the active point: \beq
\dot{w}_0=-w_0\{1+2D(1-e^{-\lam})\}+w_c+G+1/2+2D(G+w_c)(1-e^{-\lam}).
\label{eq:w0_eqn} \eeq

\subsection{$K=1$ kinetic relation}

To see whether the ensuing dynamics is compatible with the exact kinetic
relation  \eqref{eq:G_OD} in the Appendix, we solve \eqref{eq:w0_eqn} subject to
the boundary condition $w_0(0)=w_c$, obtaining \beq
w_0(t)=w_c+(G+G_{\rm P})\biggl\{1-\exp\biggl(-\dfrac{t}{2G_{\rm
P}}\biggr)\biggr\}, \label{eq:w0_soln} \eeq Then, setting  $n=1$
in \eqref{eq:other_strains_1D}, we find
\[
w_1(t)=w_c+G-G_{\rm P}-e^{-\lam}(G+G_{\rm
P})\exp\biggl(-\dfrac{t}{2G_{\rm P}}\biggr).
\]
The second boundary condition $w_1(1/V)=w_c$  yields the desired
approximation of the kinetic relation:
\beq G(V)=G_{\rm
P}+\dfrac{(1-e^{-\lam})(G_{\rm S}-G_{\rm
P})}{\exp(\frac{1}{2G_{\rm P}V})-\exp(-\lam)}.
\label{eq:G_reduced_1D} \eeq Clearly, $G(V) \rightarrow G_{\rm P}$
as $V \rightarrow 0$, and in agreement with
\eqref{eq:G_OD} $G(V)$ tends to the spinodal value
$G_{\rm S}=1/2$ as $V$ goes to infinity. Note, however, that the asymptotic behavior at
small $V$, $G-G_{\rm P} \sim \exp(-\frac{1}{2G_{\rm P}V})$,
differs from the exact asymptotics \beq G -
G_{\rm P} \approx \dfrac{\sqrt{V}\exp(-1/V)}{\sqrt{4\pi D}}.
\label{eq:smallV_approx} \eeq
obtained in \cite{Fath98}.

\subsection{$K=1$ dynamic PN landscape}

To construct the dynamic PN landscape we need to to patch different segments of the dynamic
trajectory into one by introducing a single order parameter.
Recall that during each time period $k/V \leq t \leq (k+1)/V$ the
active strain is $w_k(t)$, and the corresponding solution can be
obtained by replacing $n$ in the expressions above for $k=0$ by
$n-k$ and $t$ by $t-k/V$. Introduce a new continuous monotonically
increasing function $\nu(t)$: \beq
\begin{split}
\nu(t)=&[Vt]+\dfrac{1-\exp(-\frac{1}{2G_{\rm P}}(t-\frac{[Vt]}{V}))}{1-\exp(-\frac{1}{2G_{\rm P}})}\\
      =&[Vt]+\dfrac{G+G_{\rm P}}{(G_{\rm S}-G)(e^{\lam}-1)}
      \biggl\{1-\exp\biggl(-\frac{1}{2G_{\rm P}}\biggl(t-\frac{[Vt]}{V}\biggr)\biggr)\biggr\},
\end{split}
\label{eq:nu_1D} \eeq where $[\nu]=[Vt]$  and
note that $\nu(t)$ takes consecutive integer values at the
beginning of each time period. Therefore, it is exactly the dynamic extension
of the parameter $\nu$ which we used to obtain the static PN
landscape.
\begin{figure}
\centerline{\psfig{figure=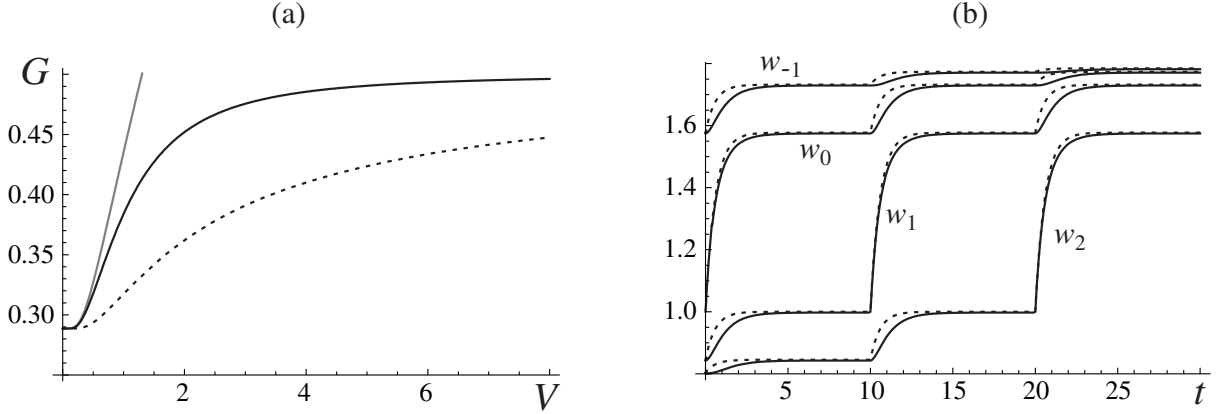,width=\textwidth}} \caption{(a)
Kinetic relations $G(V)$ for infinite-dimensional dynamics \eqref{eq:G_OD} (solid
curve), its small-velocity approximation \eqref{eq:smallV_approx}
(grey curve) and $K=1$ approximation (dotted curve). (b)
Comparison of strain trajectories obtained in the $K=1$ reduced model
(dotted curves) and in the original infinite-dimensional  model
(solid curves). Parameters: $V=0.1$, $D=0.5$, $w_c=1$.}
\label{fig:1Dreduction}
\end{figure}
In terms of this new ``global'' order parameter, we obtain the
following expression for an active strain $w_{[\nu]}(t)$ (over
the time period $[\nu]/V \leq t \leq ([\nu]+1)/V$):
\[
w_{[\nu]}(t)=w_c+(e^{\lam}-1)(G_{\rm S}-G)(\nu(t)-[\nu(t)]).
\]
Time evolution of all strains can be then written in terms of
$\nu$ as\footnote{Although $\nu(t)$ is by
construction a continuous function of time, the strain variables
have jump discontinuities at $t=n/V$, $n=1,2\dots$. This is due to the
fact that our approximation treats non-active strains as if they
were equilibrated and neglects their real dynamics. As a
result, at the end of each time period when we switch to the new active
strain, the old one has to increase its value discontinuously in order to reach equilibrium with the new active point. The associated jumps can
be computed explicitly:
$
\lbr w_n \rbr_{t=(n+1)/V}=\lbr w_0
\rbr_{t=1/V}=w_0(1/V+0)-w_0(1/V-0) =2\sinh\lam(G-G_{\rm P}).
$}
\beq
\begin{split}
&w_n(t)=w_c-1/2+G+\\
        &\left\{\begin{array}{ll}
     1+\biggl((e^{\lam}-1)(G_{\rm S}-G)(\nu(t)-[Vt])
                                         -G-G_{\rm S}\biggr)e^{\lam (n-[Vt])}, & n \leq [Vt]\\
       \biggl((e^{\lam}-1)(G_{\rm S}-G)(\nu(t)-[Vt])
                                         -G+G_{\rm S}\biggr)e^{-\lam(n-[Vt])} & n \geq [Vt]+1.
                                        \end{array}
                            \right.
\end{split}
\label{eq:all_strains_1D} \eeq
Now we are in a position to construct the global DPN potential
$\Phi(\nu;G)$ which serves as a landscape for the overdamped
dynamics of $\nu(t)$: \beq \dot{\nu}=-\alpha \Phi'(\nu;G).
\label{eq:Phi_1D1} \eeq The potential $\Phi(\nu;G)$ is piecewise
quadratic and we can select the coefficient $\alpha=1/(4G_P^2)$  to ensure that the
function \eqref{eq:nu_1D} is the solution of \eqref{eq:Phi_1D1}.
As the result, we obtain \beq \Phi(\nu;G)=G_{\rm
P}\biggl((\nu-[\nu])^2-\dfrac{2(G+G_{\rm P})}{(e^{\lam}-1)(G_{\rm
S}-G)}\nu+[\nu]\biggr). \label{eq:Phi_1D} \eeq Notice that this potential coincides with the
equilibrium PN landscape at $G=G_{\rm P}$ (see
Figure~\ref{fig:1Dpotential}).

In
Figure~\ref{fig:1Dreduction}b we compare the one-dimensional dynamics \eqref{eq:Phi_1D1} with the
full infinite-dimensional dynamics at $G=\text{const}$ (see Appendix).  As expected, the evolution of the
active strain is captured quite well, but there is a visible
deviation from the actual values for the strains whose adjustment
was assumed to be instantaneous. The corresponding kinetic
relations are compared in Figure~\ref{fig:1Dreduction}a. One can see that the reduced model with $K=1$ provides a good
quantitative approximation of the exact \emph{kinetic relation} until
$V=0.2$. At higher velocities it deviates substantially from the
exact relation, although both tend to the same spinodal limit
$G_{\rm S}$ at infinite $V$. The observed discrepancies in the case $G=\text{const}$
suggest that the reduced model with $K=1$ is oversimplified and cannot be used as the base for the construction of \emph{kinetic equations}.

\section{Reduced model with two internal variables}

To obtain a reduced model which works well in the whole range of admissible velocities, we  need to consider more ``active
points" in the core region of the defect. From
Figure~\ref{fig:1Dreduction}b one can see that in each period the
natural choice for the expanded set of order parameters would be
$w_{-1}$, $w_0$ and $w_{1}$.

\subsection{$K=2$ model}

Assume again that $G=\text{const}$ and minimize the energy with respect to all strain
variables other than $w_{-1}$, $w_0$ and $w_{1}$.\footnote{Notice that instead of minimizing out the ``enslaved'' variables at each value of $G$ we could have also used the tails of the known
traveling wave solution (see Appendix). However this leads to very complicated implicit algebraic relations and in the interest of transparency we decided not to pursue this more rigorous approach in this paper.} We obtain the following recovery relations \beq w_n=\left\{\begin{array}{ll}
             w_c+G+1/2+(w_{-1}-w_c-G-1/2)e^{\lam (n+1)}, & n \leq -1\\
             w_0, & n=0\\
             w_c+G-1/2+(w_1-w_c-G+1/2)e^{\lam(1-n)} & n \geq 1.
                                        \end{array}
                             \right.
\label{eq:other_strains_2D}\eeq
Observe that the dynamics of the variables
$w_{-1}(t)$, $w_0(t)$ and $w_1(t)$ in \eqref{eq:other_strains_2D} is not independent.  Indeed,
the three active points satisfy the dynamic equations \beq
\begin{split}
\dot{w}_{-1}&=D(w_0-2w_{-1}+w_{-2})-w_{-1}+w_c+G+1/2\\
\dot{w}_0&=D(w_1-2w_0+w_{-1})-w_0+w_c+G+1/2\\
\dot{w}_1&=D(w_2-2w_1+w_0)-w_1+w_c+G-1/2
\end{split}
\label{eq:3pt_dyn1} \eeq Substituting the expressions for $w_2$
and $w_{-2}$ from \eqref{eq:other_strains_2D} in
\eqref{eq:3pt_dyn1}, we obtain \beq
\begin{split}
\dot{w}_{-1}&=(-2D-1+e^{-\lam}D)w_{-1}+Dw_0+(w_c+G+1/2)(1+D(1-e^{-\lam}))\\
\dot{w}_0&=(-2D-1)w_0+Dw_1+Dw_{-1}+w_c+G+1/2\\
\dot{w}_1&=(-2D-1+e^{-\lam}D)w_{-1}+Dw_0+(w_c+G-1/2)(1+D(1-e^{-\lam}))
\end{split}
\label{eq:3pt_dyn2} \eeq If we now compare the equations governing
the dynamics of $w_{-1}(t)$ and $w_1(t)$, we see that they differ
only by a constant term in the right hand side. This allows us to
reduce \eqref{eq:3pt_dyn2} to a two-dimensional system for \beq
x(t)=w_0(t), \quad y(t)=\dfrac{w_{-1}(t)+w_1(t)}{2}.
\label{eq:def_of_xy} \eeq 
Note that
\[
w_{\mp1}(t)=y(t)\pm\dfrac{1+D(1-e^{-\lam})}{2(1+D(2-e^{-\lam}))}=y(t)\pm\dfrac{1-e^{-\lam}}{2},
\]
where we used \eqref{eq:lam} to obtain the second equality.
The two variables: $x(t)$, describing the dynamics of the
transforming spring, and $y(t)$, describing the average
strain in the core region, must satisfy the
following system of equations:
\beq
\dot{x}=(-2D-1)x+2Dy+w_c+G+1/2 \label{eq:x_eqn} \eeq \beq
\dot{y}=Dx+(-2D-1+e^{-\lam}D)y+(w_c+G)(1+D(1-e^{-\lam})).
\label{eq:y_eqn} \eeq
This system will be used as an
approximation of the full infinite-dimensional
dynamics.

\subsection{$K=2$ kinetic relation}

We first check that the traveling wave solutions of the original system
(see Appendix) can be reproduced by the reduced system \eqref{eq:x_eqn},
\eqref{eq:y_eqn}. Such solution must be subjected to the following constraints.
First, we must require that \beq x(0)=w_c, \label{eq:cond1} \eeq a
condition that ensures that the $0$th NN spring (recall that
$x=w_0$) has just transformed to the new phase at $t=0$. Second,
we require that
\[
w_1(1/V)=w_c,
\]
or \beq y(1/V)=w_c+\dfrac{1-e^{-\lam}}{2}. \label{eq:cond2} \eeq
This means that at $t=1/V$ the first NN spring reaches the critical
strain, marking the end of the period. Finally, periodicity
requires that
\[
w_1(0)=w_2(1/V).
\]
\begin{figure}
\centerline{\psfig{figure=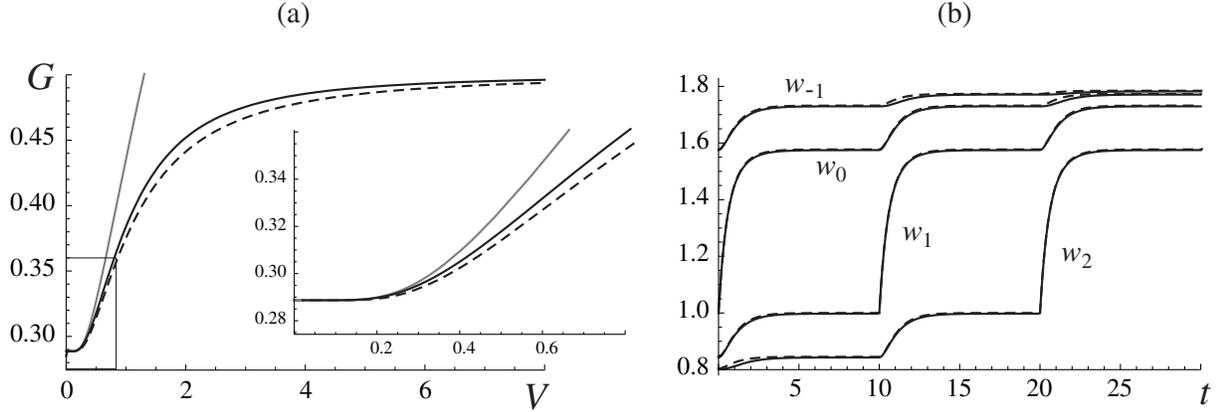,width=\textwidth}} \caption{(a)
Kinetic relations $G(V)$: infinite-dimensional system \eqref{eq:G_OD} (solid
curve), its small-velocity approximation \eqref{eq:smallV_approx}
(grey curve) and  $K=2$ approximation (dashed curve); (b)
Comparison of the strain trajectories obtained in the $K=2$ model (dashed
curves) and in the infinite-dimensional model (solid
curves). Parameters: $V=0.1$, $D=0.5$, $w_c=1$.}
\label{fig:2Dreduction}
\end{figure}
Using \eqref{eq:other_strains_2D} at $n=2$ together
with \eqref{eq:cond2}, we can see that this boundary condition
reduces to
\[
w_1(0)=w_c-(G_{\rm S}-G)(1-e^{-\lam}),
\]
or
\beq y(0)=w_c+G(1-e^{-\lam}).
\label{eq:cond3}
\eeq
Next, by solving the system \eqref{eq:x_eqn}, \eqref{eq:y_eqn}
subject to the initial conditions \eqref{eq:cond1} and
\eqref{eq:cond3}, we obtain \beq
\begin{split}
x(t)&=w_c+G+G_{\rm P}+c_1e^{r_1 t}+c_2e^{r_2 t}\\
y(t)&=w_c+G+e^{-\lam}G_{\rm
P}+\dfrac{e^{-\lam}}{4}(c_1(1+\sqrt{1+8e^{2\lam}})e^{r_1 t}
+c_2(1-\sqrt{1+8e^{2\lam}})e^{r_2 t}),
\end{split}
\label{eq:x_and_y_soln}
\eeq
where
\beq
r_{1,2}=-\dfrac{e^{\lam}}{4\sinh^2(\lam/2)}\biggl(1+\dfrac{1}{2}e^{-2\lam}(1\mp\sqrt{1+8e^{2\lam}})\biggr)
\label{eq:rates} \eeq (note that $r_2<r_1<0$) and
\[
\begin{split}
c_1&=-\dfrac{(3+\sqrt{1+8e^{2\lam}})(2G-1+e^\lam(1+2G))}{4(1+e^{\lam})\sqrt{1+8e^{2\lam}}}\\
c_2&=-\dfrac{(\sqrt{1+8e^{2\lam}}-3)(2G-1+e^\lam(1+2G))}{4(1+e^{\lam})\sqrt{1+8e^{2\lam}}}.
\end{split}
\]
The application of the boundary condition \eqref{eq:cond2} yields the desired approximation of the
exact kinetic relation:
\beq G(V)=\dfrac{G_{\rm
P}+\frac{1-e^{-\lam}}{4(1+e^\lam)\sqrt{1+8e^{2\lam}}}
\biggl((1+2e^{2\lam}+\sqrt{1+8e^{2\lam}})e^{r_1/V}+(\sqrt{1+8e^{2\lam}}-2e^{2\lam}-1)e^{r_2/V}\biggr)}
{1-\frac{e^{-\lam}}{2\sqrt{1+8e^{2\lam}}}
\biggl((1+2e^{2\lam}+\sqrt{1+8e^{2\lam}})e^{r_1/V}+(\sqrt{1+8e^{2\lam}}-2e^{2\lam}-1)e^{r_2/V}\biggr)}.
\label{eq:G_reduced_2D} \eeq

\begin{figure}
\centerline{\psfig{figure=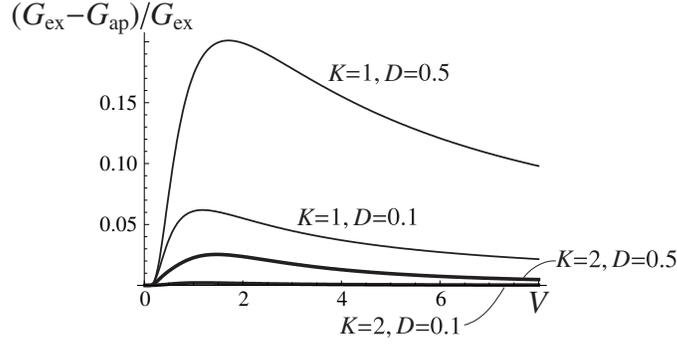,width=3.5in}} \caption{The
error in the approximation of the the exact kinetic relation
$G_{\rm ex}(V)$ by the reduced models with $K=1$ and $K=2$
generating approximate kinetic relation $G_{\rm ap}(V)$.}
\label{fig:Gerror}
\end{figure}

First we notice that the approximate kinetic relation \eqref
{eq:G_reduced_2D} satisfies the constraint $G(0)=G_{\rm P}$. Then,
as $V$ tends to infinity, $G(V) \rightarrow G_{\rm S}=1/2$ with
the asymptotics  $G-G_{\rm s} \sim \exp(-|r_1|/V)$. The global comparison
of the approximate and exact (see Appendix) kinetic relations is presented in
Figure~\ref{fig:2Dreduction}.  One can see that in view of how few
degrees of freedom are involved in the approximation, the
agreement is remarkable, at least for small (strongly discrete limit) to moderate values of
$D$ (see Figure~\ref{fig:Gerror}).\footnote{At larger $D$ (strong coupling limit) the core region delocalizes and more
and more active points need to be included \cite{CB2}.  Another possibility is to change the
type of the approximation from ``active points" to ``collective variables",
which is more in tune with a close to continuum character of the model in this limit.
We will not pursue these ideas in the present paper.}

\subsection{$K=2$ dynamic PN landscape}

We can now introduce the global order parameter and
reformulate the $G=\text{const}$ dynamics in the form of a gradient flow on a fixed
two-dimensional DPN landscape. Recall that expressions
\eqref{eq:x_and_y_soln}, \eqref{eq:def_of_xy} and
\eqref{eq:other_strains_2D} define the strain trajectories
$w_n(t)$ only over the time period $0 \leq t \leq 1/V$. The
extension of this solution for $k/V \leq t \leq (k+1)/V$  is
obtained by replacing $n$ in the above formulas by $n-k$ and $t$
by $t-k/V$. To patch together different periods we need to
introduce a set of two order parameters that change continuously
and monotonically with $t$. We denote these variables as
$\tilde{\nu}_1(t)$ and $\tilde{\nu}_2(t)$ and define them by the
conditions $[\tilde{\nu}_1(t)]=[\tilde{\nu}_2(t)]=[Vt]$. Then the
variables $x(t)$ and $y(t)$ can be extended periodically to any
$t\geq 0$ as follows:
$x(t)=w_c+(\tilde{\nu}_1(t)-[Vt])(x(1/V)-w_c)$ and
$y(t)=y(0)+(\tilde{\nu}_2(t)-[Vt])(y(1/V)-y(0))$. Under the
assumption that $V=\Gamma(G)$ , where $\Gamma(G)$ is the inverse of $G(V)$ from (\ref{eq:G_reduced_2D}), the dynamics of the vector field
$\boldsymbol{\tilde{\nu}}(t)=(\tilde{\nu}_1(t),\tilde{\nu}_2(t))$
can be represented in the form
\beq
\dfrac{d}{dt}\boldsymbol{\tilde{\nu}}=-\boldsymbol{\tilde{\alpha}}\nabla\tilde{\Phi}(\boldsymbol{\tilde{\nu}};G).
\label{eq:nu_11} \eeq Here we introduced a mobility matrix
$\boldsymbol{\tilde{\alpha}}$ and a two-dimensional effective DPN
potential $\tilde{\Phi}(\boldsymbol{\tilde{\nu}},G)$.
\begin{figure}
\centerline{\psfig{figure=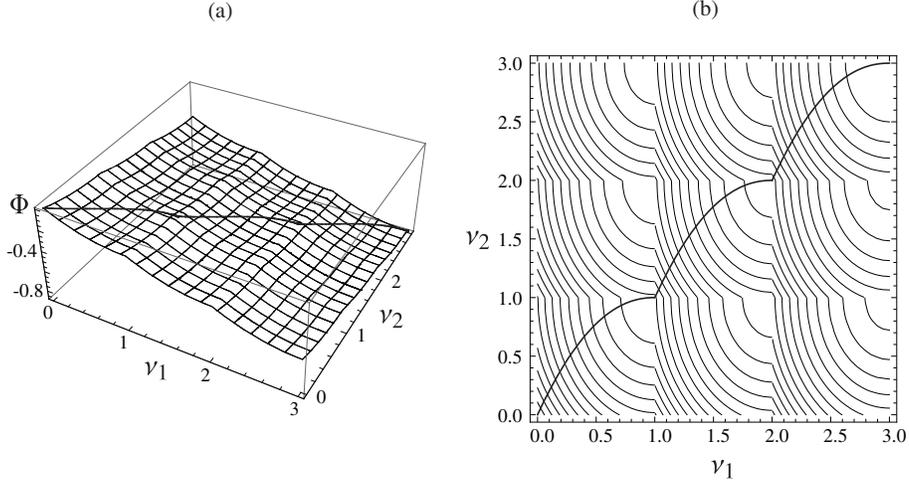,width=120mm}} \caption{(a)
Two-dimensional energy landscape $\Phi(\nu_1,\nu_2)$. (b) Level
sets of the energy
 along with the path of the effective particle $\boldsymbol{\nu}(t)$. Parameters: $V=0.1$, $D=0.5$, $w_c=1$.}
\label{fig:2Dpotential}
\end{figure}

We can further simplify the system
\eqref{eq:nu_11} by diagonalizing the mobility matrix
$\boldsymbol{\tilde{\alpha}}$. Observe that in \eqref{eq:rates} we
have $|r_2|>|r_1|>0$, so that the eigenvector
$(\frac{e^{-\lam}}{4}(\sqrt{1+8e^{2\lam}}+1),1)$ corresponding the
eigenvalue $r_2$ is the slow direction, while the eigenvector
$(\frac{e^{-\lam}}{4}(1-\sqrt{1+8e^{2\lam}}),1)$ that corresponds
to $r_1$ is the fast direction. Introduce the
\emph{slow} variable \beq
\nu_1(t)=[Vt]+\dfrac{1-\exp(r_1(t-[Vt]/V))}{1-\exp(r_1/V)}
\label{eq:nu_1} \eeq and the \emph{fast} variable \beq
\nu_2(t)=[Vt]+\dfrac{1-\exp(r_2(t-[Vt]/V))}{1-\exp(r_2/V)}.
\label{eq:nu_2} \eeq In terms of the new variables, the time
evolution of all strains is given by \beq
\begin{split}
&w_n(t)=\begin{cases}
             w_c+G+1/2+\{G_{\rm P}-1/2+\dfrac{1}{4}c_1(1+\sqrt{1+8e^{2\lam}})(1-(1-e^{r_1/V})(\nu_1(t)-[Vt]))\\
                                         \quad +c_2(1-\sqrt{1+8e^{2\lam}})(1-(1-e^{r_2/V})(\nu_2(t)-[Vt]))\}e^{\lam(n-[Vt])}, \quad n \leq [Vt]-1\\
             w_c+G+G_{\rm P}+c_1(1-(1-e^{r_1/V})(\nu_1(t)-[Vt]))\\
                                         \quad +c_2(1-(1-e^{r_2/V})(\nu_2(t)-[Vt])), \quad n = [Vt]\\
             w_c+G-1/2+\{G_{\rm P}+1/2+\dfrac{1}{4}c_1(1+\sqrt{1+8e^{2\lam}})(1-(1-e^{r_1/V})(\nu_1(t)-[Vt]))\\
                                         \quad +c_2(1-\sqrt{1+8e^{2\lam}})(1-(1-e^{r_2/V})(\nu_2(t)-[Vt]))\}e^{-\lam(n-[Vt])}, \quad n \geq [Vt]+1.
                                        \end{cases}
\end{split}
\label{eq:all_strains_2D} \eeq The vector field
$\boldsymbol{\nu}(t)=(\nu_1(t),\nu_2(t))$ satisfies
\beq
\dot{\boldsymbol{\nu}}=-\boldsymbol{\alpha}\nabla\Phi(\boldsymbol{\nu};G),
\label{eq:nu_system} \eeq where the mobility matrix
$\boldsymbol{\alpha}=\text{diag}(\alpha_1,\alpha_2)$ is now
diagonal (here $\alpha_1>0$ and $\alpha_2>0$ are constants).  The
potential $\Phi(\nu_1,\nu_2;G)$ is piecewise quadratic; it depends
on $V$ and hence, through the kinetic relation $V=\Gamma(G)$, on the driving
force $G$. Using \eqref{eq:nu_1} and \eqref{eq:nu_2}, we obtain
\[
\begin{split}
\Phi(\nu_1,\nu_2; G)=-&\dfrac{r_1}{2\alpha_1}\biggl\{(\nu_1-[\nu_1])^2-\dfrac{2}{1-\exp(r_1/V(G))}\nu_1+[\nu_1]\biggr\}\\
-&\dfrac{r_2}{2\alpha_2}\biggl\{(\nu_2-[\nu_2])^2-\dfrac{2}{1-\exp(r_2/V(G))}\nu_2+[\nu_2]\biggr\}.
\end{split}
\]

To determine the effective viscosities $\alpha_1$ and $\alpha_2$,
we require that at $G_{\rm P}$ the DPN potential equals the
relative Gibbs free energy of the system:
$$\Phi(\boldsymbol{\nu};G_{\rm P})=\mathcal{W}(\boldsymbol{\nu};G_{\rm P})-\mathcal{W}(\mathbf{0};G_{\rm P}).$$
This yields \beq
\alpha_1=-\dfrac{r_1\sqrt{1+8e^{2\lam}}}{(\sqrt{1+8e^{2\lam}}+3)G_{\rm
P}}, \quad
\alpha_2=-\dfrac{r_1\sqrt{1+8e^{2\lam}}}{(\sqrt{1+8e^{2\lam}}-3)G_{\rm
P}} \label{eq:alphas} \eeq and, finally, \beq
\begin{split}
\Phi(\nu_1,\nu_2; G)=&\dfrac{(\sqrt{1+8e^{2\lam}}+3)G_{\rm
P}}{2\sqrt{1+8e^{2\lam}}}
\biggl\{(\nu_1-[\nu_1])^2-\dfrac{2}{1-\exp(r_1/V(G))}\nu_1+[\nu_1]\biggr\}\\
+&\dfrac{(\sqrt{1+8e^{2\lam}}-3)G_{\rm P}}{2\sqrt{1+8e^{2\lam}}}
\biggl\{(\nu_2-[\nu_2])^2-\dfrac{2}{1-\exp(r_2/V(G))}\nu_2+[\nu_2]\biggr\}.
\end{split}
\label{eq:Phi_2D} \eeq The dynamic PN landscape (DPN)  $\Phi(\nu_1,\nu_2; G)$,
along with the trajectory of the effective particle, is shown in
Figure~\ref{fig:2Dpotential}. The comparison of strain
trajectories in the two-dimensional reduced theory with the exact
result at $G=\text{const}$ (see Appendix) shows the considerable improvement over the one-dimensional
approximation. For instance, Figure~\ref{fig:2Dreduction}b
compares the evolution of strains near the phase boundary over
the first three time periods at $V=0.1$. While in
the case $K=1$ the evolution of only the transforming element is
followed closely over each time period, in the $K=2$ case
the dynamics of the main nontransforming elements forming the core region of the defect is also captured
extremely well. It is then natural to use the $K=2$ approximation as the basis for the construction of the nonlocal rheological relation on the phase boundary.

\section{Kinetic equations}

In the previous section we saw that the traveling wave
solution of the original infinite-dimensional system can be
reproduced rather faithfully by the reduced two-dimensional system
\eqref{eq:x_eqn} and \eqref{eq:y_eqn}. This suggests that the $K=2$ model
may be extended to deal with the situation when the driving
force is not constant but varies sufficiently slowly. In this more
general setting we obtain the system \beq
\begin{split}
\dot{x}=&(-2D-1)x+2Dy+w_c+G(t)+1/2\\
\dot{y}=&Dx+(-2D-1+e^{-\lam}D)y+(w_c+G(t))(1+D(1-e^{-\lam})).
\end{split}
\label{eq:xy_system} \eeq
Equations \eqref{eq:xy_system} can be interpreted as the system of
differential \emph{kinetic equations} generalizing the \emph{kinetic relation} \eqref{eq:G_reduced_2D}.
According to the description of kinetics implied by the system \eqref{eq:xy_system}
the adjustment of the velocity $V(t)$ to the dynamic configurational loading $G(t)$ is
not instantaneous and cannot be described by an algebraic
relation if the function $G(t)$ changes sufficiently fast. Instead, the relationship
between the velocity and the driving force is expected to be history-dependent:
\beq V_*(t)=\mathcal{F}\{G(\tau),\tau
\leq t\}. \label{eq:nonlocal_rel} \eeq The goal of this section is
to reconstruct the nonlocal relationship \eqref{eq:nonlocal_rel} from  the system
\eqref{eq:xy_system} and to illustrate the effect of nonlocality.

Suppose that the phase boundary is located at $n=i$ at  time
$t=t_i$, meaning that in this instance the $i$th spring has
critical strain: $w_i(t_i)=w_c$. As before, we can define
$x(t)=w_i(t)$ and $y(t)=(w_{i+1}(t)+w_{i-1}(t))/2$ when $t_i \leq
t \leq t_{i+1}$. To obtain the average velocity $\bar{V_i}$ we
need to solve the system of kinetic equations \eqref{eq:xy_system}
in the interval $(t_i,t_{i+1})$, under the assumption that $G(t)$ is known.

Observe first that conditions \beq
x(t_i)=w_c \label{eq:C1} \eeq and \beq
y(t_{i+1})=w_c+\dfrac{1-e^{-\lam}}{2}, \label{eq:C2} \eeq which
are the extensions of \eqref{eq:cond1} and \eqref{eq:cond2} to the
interval $[t_i,t_{i+1}]$, ensure that the system holds from
$t=t_i$ until the time $t_{i+1}$ when the next spring reaches the
critical strain: $w_{i+1}(t_{i+1})=w_c$.

While for given $t_{i+1}$ conditions \eqref{eq:C1} and
\eqref{eq:C2} are sufficient to find a unique solution of
\eqref{eq:xy_system}, the  moment
 $t_{i+1}$ when the consecutive spring switches from one energy well to another remains unknown.
 If $G(t)$ varies sufficiently slowly, it is reasonable to assume that
the motion is close to periodic in the sense that\footnote{This condition is
obviously not exact because  $G(t_i)\neq G(t_{i+1})$. By using the
knowledge of the exact structure of the traveling waves at both
$G(t_i)$ and $G(t_{i+1})$ this approximation can be in principle improved. However, in the interests of transparency, we do not pursue this more rigorous approach in this paper.} \beq
w_i(t_i)=w_{i+1}(t_{i+1}). \label{eq:CCCC} \eeq  This condition allows us to close the system and find $t_{i+1}$ from
 \beq y(t_i)=w_c+G(t_{i+1})(1-e^{-\lam}).
\label{eq:C3} \eeq
To this end, we solve
\eqref{eq:xy_system} subject to \eqref{eq:C1} and \eqref{eq:C3},
 obtain  for $t_i \leq t \leq t_{i+1}$, \beq
\begin{split}
x(t)=&g(t)+\int_{t_i}^{t}(e^{r_1(t-\tau)}f_1(\tau)+e^{r_2(t-\tau)}f_2(\tau))d\tau\\
y(t)=&h(t)+\dfrac{e^{-\lam}}{4}\int_{t_i}^{t}\{(1+\sqrt{1+8e^{2\lam}})e^{r_1(t-\tau)}f_1(\tau)\\
                                           &+(1-\sqrt{1+8e^{2\lam}})e^{r_2(t-\tau)}f_2(\tau)\}d\tau,
\end{split}
\label{eq:wixi} \eeq where we defined \beq
\begin{split}
g(t)=&\dfrac{1}{\sqrt{1+8e^{2\lam}}} \biggl\{\dfrac{1}{2}\biggl(
(\sqrt{1+8e^{2\lam}}-1)e^{r_1(t-t_i)}+(\sqrt{1+8e^{2\lam}}+1)e^{r_2(t-t_i)}
\biggr)w_c\\
+&2e^\lam(e^{r_1(t-t_i)}-e^{r_2(t-t_i)})(w_c+G(t_{i+1})(1-e^{-\lam}))
\biggr\}\\
h(t)=&\dfrac{1}{\sqrt{1+8e^{2\lam}}}
\biggl\{e^\lam(e^{r_1(t-t_i)}-e^{r_2(t-t_i)})w_c\\
&+\dfrac{1}{2}\biggr((\sqrt{1+8e^{2\lam}}+1)e^{r_1(t-t_i)}+(\sqrt{1+8e^{2\lam}}-1)e^{r_2(t-t_i)}\biggr)(w_c+G(t_{i+1})(1-e^{-\lam}))\biggr\}\\
f_1(t)=&\dfrac{\sqrt{1+8e^{2\lam}}-1}{2\sqrt{1+8e^{2\lam}}}(w_c+\dfrac{1}{2}+G(t))
       +\dfrac{2e^\lam}{\sqrt{1+8e^2\lam}}(w_c+G(t))(1+D(1-e^{-\lam}))\\
f_2(t)=&\dfrac{\sqrt{1+8e^{2\lam}}+1}{2\sqrt{1+8e^{2\lam}}}(w_c+\dfrac{1}{2}+G(t))
       -\dfrac{2e^\lam}{\sqrt{1+8e^2\lam}}(w_c+G(t))(1+D(1-e^{-\lam}).
\end{split}
\label{eq:f1f2} \eeq
We can now determine the strains $w_i(t)=x(t)$ and $w_{i\mp1}(t)=y(t)\pm(1-e^{-\lam})/2$. The other strains can be recovered as
before, from the constrained energy minimization:
\[
w_n(t)=\left\{\begin{array}{ll}
             w_c+G(t)+1/2+(w_{i-1}-w_c-G(t)-1/2)e^{\lam (n-i+1)}, & n \leq i-2\\
             w_c+G(t)-1/2+(w_{i+1}-w_c-G(t)+1/2)e^{\lam(1-n-i)} & n \geq i+2.
                                        \end{array}
                             \right.
\]
Condition \eqref{eq:C2} applied to $y(t)$ in
\eqref{eq:wixi} yields the nonlinear equation for $t_{i+1}$: \beq
\begin{split}
h(t_{i+1})-w_c-\dfrac{1-e^{-\lam}}{2}+
\dfrac{e^{-\lam}}{4}\int_{t_i}^{t_{i+1}}&\{(1+\sqrt{1+8e^{2\lam}})e^{r_1(t_{i+1}-\tau)}f_1(\tau)\\
                                           +&(1-\sqrt{1+8e^{2\lam}})e^{r_2(t_{i+1}-\tau)}f_2(\tau)\}d\tau=0.
\end{split}
\label{eq:tip1}
\eeq
As we have seen in the previous section, at $G=\text{const}$  this equation a unique solution and one
can expect that at least for the case of slowly varying driving force $G(t)$
 the solution $t_{i+1}$  also exists and is unique. Once it is found, we can
obtain the average velocity corresponding to the time interval $[t_i,t_{i+1}]$ in the form
\beq
\bar{V_i}=\dfrac{1}{t_{i+1}-t_i}
\label{eq:V_1}
\eeq
Using this iterative  process, one can generate the sequence of successive time intervals
$[t_i,t_{i+1}]$ and find the average velocities associated with each of them. This gives a piecewise constant function $\bar V(t_i)$,
which appears at the macroscale as a  continuous function $V_*(t)$.
\begin{figure}
\centerline{\psfig{figure=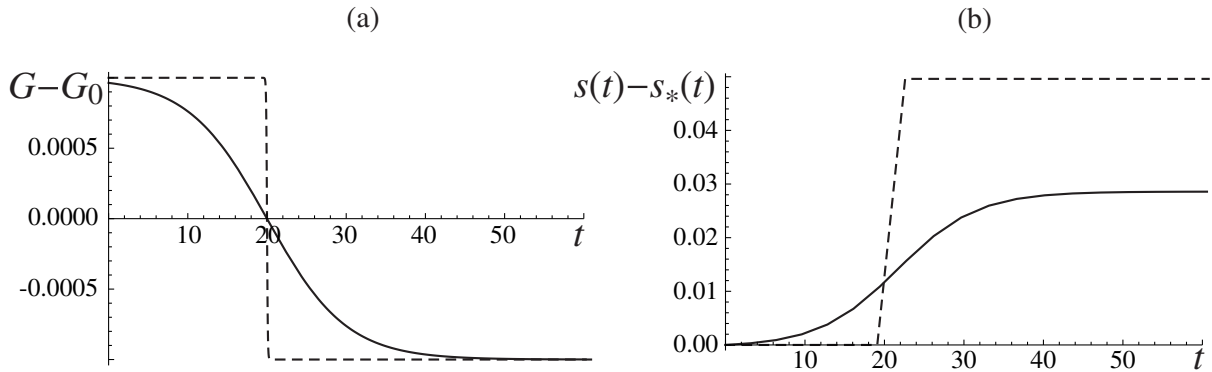,width=\textwidth}}
\caption{(a) Time-dependent driving force \eqref{eq:example}, with $\zeta=0.1$
(dashed curve) and $\zeta=10$ (solid curve). (b) The corresponding
difference between the coordinate $s(t)$ of the phase boundary
obtained using the instantaneous kinetic response $V(t)=\Gamma(G(t))$ and
the coordinate $s_*(t)$ calculated using the kinetic equations
\eqref{eq:xy_system}. Here $G_0=G(0.3) \approx 0.294$, $\Gamma=0.001$ and $t_*=20$.} \label{fig:example}
\end{figure}

To illustrate this procedure consider a specific time-dependent driving force of the form
\beq
G(t)=G_0-\delta\tanh\dfrac{t-t_*}{\zeta}.
\label{eq:example}
\eeq
Here $G_0=G(t_*)$ is the constant background driving force,
$\delta$ is the amplitude of the diffuse jump in the driving force, and $\zeta$ is the characteristic time over which
the driving force decreases by $2\delta$. Such
perturbation of the driving force may represent an interaction of the phase boundary
with an extended obstacle.

We now compare two responses of the steadily moving phase boundary to a perturbation \eqref{eq:example}.
One is the instantaneous response governed by the  kinetic
relation \eqref{eq:G_reduced_2D}. In this case we obtain $V(t)=\Gamma(G(t))$ and the location of the phase boundary assuming $s(0)=0$ can be found by integration from
\[
s(t)=\int_{0}^t \Gamma(G(\tau))d\tau.
\]
The second one is the history-dependent
response governed by the kinetic equations \eqref{eq:xy_system}. 
In this case the function $V_*(t)$ is obtained
from the interpolation of the solution of the sequence of the
discrete problems \eqref{eq:tip1}. The location of the phase boundary is given by
\[
s_{*}(t_i)=\sum_{j=0}^{i} \bar{V}_j (t_j-t_{j-1})=i
\]
The time-dependent separation between the two boundaries both located at the same point at time $t=0$ but then driven by two
 different kinetic models is shown in Fig.~\ref{fig:example}. As $\zeta\rightarrow\infty$, the two kinetic models converge and 
the difference $s(t)-s_*(t)$ tends to zero.  However, as Fig.~\ref{fig:example} shows, at finite $\zeta$ the prediction of the model 
based on the \emph{kinetic relation} may be markedly different from the one based on solving the \emph{kinetic equations}.

\section{Conclusions}

In the classical macroscopic continuum description of lattice defects the core
regions are represented by singularities, and the dynamics of these singularities is governed by algebraic kinetic relations.
Such description is too coarse to capture the details of the dynamic response of the cores of the defects to relatively fast changes of the macroscopic
driving forces. In particular, it averages out the details of the intricate interaction of the core regions with localized
inhomogeneities which manifests itself through macroscopic velocity oscillations and a characteristic acoustic emission.

To describe the ``breathing" of the core regions as well as other transient effects that are usually neglected,
we proposed in this paper to replace the algebraic relations between the macroscopic velocity and the corresponding
driving force, which imply internal steady state, by a system of differential kinetic equations.
The explicit time dependence of the driving force makes these equations  nonautonomous
and allows them to represent a dependence between the current value of velocity
and the history of the driving force. In the constitutive sense the proposed kinetic
equations describe a relation between the thermodynamic force and
the corresponding flux which is nonlocal in time. The ensuing rheological description of the moving defect can be qualified as a differential rate model.

The main idea of our construction is to follow the exact dynamics of only few discrete degrees
of freedom. The adiabatic elimination of the infinite number of the remaining variables outside the core region serves 
as the matching condition with the classical continuum description outside the defect. The choice of the dimensionality 
of the reduced system remains heuristic and is based on the comparison of some special periodic solutions of the finite-dimensional system with the
discrete traveling wave solutions of the microscopic infinite-dimensional model.
We show that in contrast to the standard approach that involves only the center of mass of the defect, the minimally adequate approximation includes
not only the location but also the internal configuration of the
core. Such extension of the scope is necessary even if both the macroscopic driving force and the average velocity of the defect are constant 
because the core region may experience
periodic configurational deformations instead of
translating as a rigid body.

In this paper we presented only the contours of this approach and provided only the most elementary illustrations.
More systematic study is needed to formulate the formal mathematical structure of our center-manifold-type reduction
procedure and to evaluate the errors of the approximation in mathematical terms. Other problems include transition from
overdamped to underdamped dynamics and the generalization for higher dimensions. A specific development is also needed
to move from a prototypical case of martensitic phase transitions to the cases of dislocations, cracks, and more complex singularities.

\section*{Appendix: Traveling wave solution}

To construct a traveling wave solution of the system
\eqref{eq:main} with $\sigma=\text{const}$, we assume that \beq
w_n(t)=w(\eta),\;\;\;\;\eta=n-Vt, \label{eq:ansatz} \eeq where $V$
is the dimensionless velocity of the front, which represents a
moving phase boundary. Since we seek a description of an
isolated  phase boundary that leaves phase II behind, we require that \beq
w(\eta)<w_c\;\;\;\;\mbox{\rm for $\eta>0$},\;\;\;\;\;\;
w(\eta)>w_c\;\;\;\; \mbox{\rm for $\eta<0$} \label{eq:constraints}
\eeq Under these assumptions, (\ref{eq:main}) reduces to \beq
Vw'(\eta)-D(w(\eta+1)-2w(\eta)+w(\eta-1))+w(\eta)=\theta(-\eta)+\sigma
\label{eq:TW} \eeq At infinity the solution must tend to
uniform-strain equilibria of (\ref{eq:rescaled_OD}): \beq w(\eta)
\rightarrow w_{\pm}\;\;\;\; \mbox{\rm as $\eta \rightarrow \pm
\infty$}. \label{eq:wpm} \eeq Finally, for consistency, we must
also require that \beq w(0)=w_c. \label{eq:switch} \eeq

\begin{figure}
\centerline{\psfig{figure=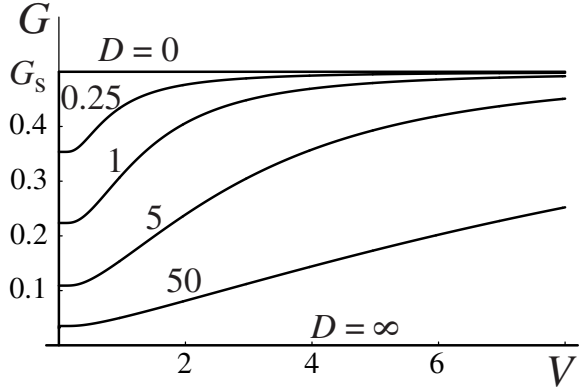, width=3.in}} \caption{Kinetic
relation at different values of $D$.} \label{fig:GvsV_OD}
\end{figure}

Since the equation (\ref{eq:TW}) is linear, we can be solve it
using Fourier transform (see \cite{CB2,Fath98,TV5} for details).
We obtain \beq w(\eta)=\left\{ \begin{array}{ll}
                    \sigma+1+\underset{k \in S^-(V)}\sum
\dfrac{e^{ik \eta}}{k \Lambda_k(k,V)}
& \mbox{for $\eta<0$} \\
                    \sigma-\underset{k \in S^+(V)}\sum
\dfrac{e^{ik \eta}}{k \Lambda_k(k,V)} & \mbox{for $\eta>0$},
                    \end{array}
            \right.
                   \label{eq:TW_soln_OD}
\eeq where $S^{\pm}(V)=\{k:\Lambda(k,V)=0,\;{\rm Im k}\gtrless
0\}$ are the sets of roots of the dispersion relation
\[
\Lambda(k,V) \equiv 1+4D\sin^2(k/2)-Vik=0.
\]
Continuity of $w(\eta)$ at $\eta=0$ gives the relationship between
the applied stress and velocity of the traveling wave: \beq
\sigma=\sigma_M+\dfrac{1}{2} +\underset{k \in
S^+(V)}\sum\dfrac{1}{k \Lambda_k(k,V)}
=\sigma_M-\dfrac{1}{2}-\underset{k \in S^-(V)}\sum \dfrac{1}{k
\Lambda_k(k,V)}. \label{eq:sigma_vs_V} \eeq  Since the
difference between applied and Maxwell stresses is equal to the
driving force $G=\sigma-\sigma_{\rm M}$ (see \cite{TV3a}),  we
obtain\footnote{This semianalytic expression remains implicit until the exact locations of the roots $k=k(V)$ of the dispersion equation are known \cite{TV5}.}
 \beq G(V)= \dfrac{1}{2}
+\underset{k \in S^+(V)}\sum\dfrac{1}{k \Lambda_k(k,V)}.
\label{eq:G_OD} \eeq
The structure of the kinetic relation (\ref{eq:G_OD}) is illustrated in
Figure~\ref{fig:GvsV_OD} for
different values of $D$.
One can see that at $D=0$ (no NNN interactions) the driving force
 must be constant and equal to the
spinodal value $G=G_{\rm S}=1/2$; at $V=0$ it can take any value
between $0$ and $G_{\rm S}$. In the limit  $D \rightarrow \infty$ the
driving force becomes equal to zero at all values of $V$. At large $V$ complex roots in the set $S^+(V)$ tend to
infinity  and the kinetic curves approach the common
limit. For
$D>1/12$ the large-velocity discrete kinetics is well approximated
by the formula \cite{TV5}
\[
G(V)=\dfrac{V}{2\sqrt{V^2+4D-1/3}},
\]
which describes the kinetics of the overdamped
viscosity-capillarity model governed by the partial differential
equation $w_t=(D-1/12)w_{xx}-\hat{\sigma}(w)+\sigma$.

\bigskip

\noindent {\bf Acknowledgements.} This work was supported by the
US National Science Foundation grant DMS-0443928 (A.V.) and by the EU
contract MRTN-CT-2004-505226 (L.T.).

\end{document}